\documentstyle[12pt]{article}
\title{On ``Effective potential for a covariantly constant gauge field in 
curved spacetime''}
\author{Frank Antonsen and Karsten Bormann\\ Niels Bohr Institute \\ 
Blegdamsvej 17, DK-2100 Copenhagen \O\\ Denmark}
\date{}
\begin{document}
\maketitle
\begin{abstract}
We extend recent work by Elizalde et al. to incorporate curvatures which are
not small and backgrounds which are not just $S^2\times R^2, S^1\times S^1
\times R^2$. Some possible problems in their paper is also pointed out.
\end{abstract}

\section*{Introduction}
The calculation of effective potentials for gauge fields to one loop order
and beyond is certainly an important task. Recently Elizalde, Odintsov and 
Romeo, \cite{eli}, has carried out such a calculation for a covariantly 
constant $SU(N)$-field on a curved background of the form $S^2\times R^2,
S^1\times S^1\times R^2$, in the limit of small curvature (i.e. large radii).
Elsewhere we have considered the same problem in a more general setting, 
\cite{os}, namely a general Yang-Mills field on a curved background with
not too violently varying curvature. We believe that this method solves some
of the problems faced by Elizalde et al. A fuller investigation of this
is in the making and will be submitted shortly, \cite{os2}. \footnote{In the
first of these two papers, a mistake has crept in: a term of the form
$\delta^a_b R_m^n$, where $R_m^n$ is the Ricci tensor was missing, this 
mistake has been corrected in \cite{os2}, which also contains many 
applications.} 

Elizalde has also given a thorough discussion of the techniques underlying
his work with Odintsov and Romeo in \cite{eli2}.

As shown in \cite{os}, the effective Lagrangian for a general Yang-Mills field
on a general curved background can be written as
\begin{eqnarray}
    V_{\rm eff}(A) &=&(4\pi)^{-2}{\rm Tr}~\left(-\frac{g^6}{128}{\cal A}^2
      \ln~\frac{g^2}{4}{\cal A} 
    +\frac{3g^6}{256}{\cal A}^2\right.\nonumber\\
    &&\hspace{20mm}\left.-\frac{1}{2}\left(\ln~\frac{g^2}{4}{\cal A}\right)~{\cal B}-
      \frac{16}{3g^4}{\cal A}^{-1}{\cal C}\right)-
    \frac{g^2}{4}F_{mn}^a F^{mn}_a \label{eq:veff}
\end{eqnarray}
where $\cal A,B,C$ are matrices defined by
\begin{eqnarray}
    {\cal A}^{m(a)}_{n(b)} &=& \left(\partial_p{\cal E}^{mp}_n+\frac{3}{4}
      {\cal E}^{mp}_k 
    {\cal E}_{np}^k\right)\delta^a_b+gf_{b\hspace{3pt}c}^{\hspace{3pt}a}
      (\partial_nA^{mc}-
      \partial^mA^c_n)+\nonumber\\
    &&\qquad\frac{1}{2}\delta^m_ng^2f_{ebc}f_{d\hspace{3pt}c}^{
      \hspace{3pt}a}A^e_pA^{pd} +\delta^a_b R^m_n
    \label{eq:A}\\
    {\cal B}_{n(b)}^{m(a)} &=& \Box_0{\cal A}_{n(b)}^{m(a)} 
	\equiv \eta^{pq}\partial_p\partial_q {\cal A}_{n(b)}^{m(a)}
	\label{eq:B}\\
    {\cal C}^{m(a)}_{n(b)} &=& (\partial_p {\cal A}^{m(a)}_{k(c)})(\partial^p 
      {\cal A}^{k(c)}_{n(b)})
      \label{eq:C}
\end{eqnarray}
where
\begin{equation}
    {\cal E}_n^{mp} = \left(\partial_ne^{m\mu}-\partial^me_n^\mu\right) 
      e^p_\mu \label{eq:E}
\end{equation}
and $\partial_m = e^\mu_m\partial_\mu$. The quantities $e_m^\mu$ are vierbeins,
$g^{\mu\nu} = e^\mu_m e^\nu_n\eta^{mn}$, and $A_m^a = e^\mu_m A_\mu^a$ etc.

In the case studied in \cite{eli} with 
\begin{equation}
	A_\mu^a = -\frac{1}{2}F_{\mu\nu}^ax^\nu
\end{equation}
covariantly constant ($F_{\mu\nu}^a=const$) we get the the gauge part of these
quantities to be
\begin{eqnarray}
	{\cal A}^{m(a)}_{n(b)} &=& -\frac{1}{2}gf_{b~c}^{~a}\left(e^\mu_n 
	(\partial_\mu e^\nu_p)\eta^{mp} F^c_{\nu\rho}x^\rho -e^\nu_p
	(\partial_\nu e^\mu_n)\eta^{mp} F^c_{\mu\rho}x^\rho
	+e^\mu_n e^\nu_p\eta^{mp}F^c_{\mu\nu}\right)\nonumber\\
	&&+\frac{1}{8}g^2\delta^m_n f_{ebc}f_d^{~ac}g^{\mu\nu} F_{\mu\rho}^d
	F_{\nu\sigma}^d x^\rho x^\sigma+\mbox{curvature terms}\\
	{\cal B}^{m(a)}_{n(b)} &=&-\frac{1}{2}gf_{b~c}^{~a}
	\eta^{rs}\eta^{mp} e^\rho_r\left[
	\partial_\rho (e^\sigma_s\partial_\sigma(e^\mu_n\partial_\mu e^\nu_p))
	F_{\nu\kappa}^c x^\kappa-\partial_\rho(e^\sigma_s\partial_\sigma(
	e^\nu_p\partial_\nu e^\mu_n))F^c_{\mu\kappa}x^\kappa\right]
	\nonumber\\
	&&-\frac{1}{2}gf_{b~c}^{~c}\eta^{rs}\eta^{mp}e^\rho_r\left[
	\partial_\rho(e^\sigma_se^\mu_n\partial_\mu e^\nu_p)F_{\nu\rho}^c
	-\partial_\rho(e^\sigma_s e^\nu_p\partial_\nu e^\mu_n)F_{\mu\rho}^c
	-2\partial_\rho(e^\sigma_s\partial_\sigma(e^\mu_n e^\nu_p))
	F_{\mu\nu}^c\right]\nonumber\\
	&&+\frac{1}{8}g^2 \eta^{rs}\delta^m_n f_{ebc}f_d^{~ac}F_{\mu\lambda}^e
	F_{\nu\kappa}^d e^\rho_r\partial_\rho(e^\sigma_s\partial_\sigma(
	g^{\mu\nu}x^\lambda x^\kappa))+\mbox{curvature terms}\\
	{\cal C}^{m(a)}_{n(b)} &=& g^{\rho\sigma}\left[(\partial_\rho(e^\mu_k
	\partial_\mu e^\nu_p)\eta^{mp}A_\nu^c
	-\frac{1}{2}e^\mu_k(\partial_\mu e^\nu_k)\eta^{mp}F_{\nu\rho}^c
	\right.\nonumber\\
	&&\left.-\partial_\rho(e^\nu_p\partial_\nu e^\mu_k)\eta^{mp}A_\mu^c
	+\frac{1}{2}e^\nu_p(\partial_\nu e^\mu_k)\eta^{mp}F_{\mu\rho}^c
	+\partial_\rho(e^\mu_ke^\nu_p)\eta^{mp}F_{\mu\nu}^c\right]
	\times\left[\partial_\sigma(e^\lambda_n\partial_\lambda e^\kappa_q)
	\eta^{kq}A_\kappa^d\right.\nonumber\\
	&&\left.-\frac{1}{2}e^\lambda_n(\partial_\lambda e^\kappa_q)\eta^{kq}
	F_{\kappa\sigma}^d-\partial_\sigma(e^\kappa_q\partial_\kappa 
	e^\lambda_n)\eta^{kq}A_\lambda^d
	+\frac{1}{2}e^\kappa_q(\partial_\kappa e^\lambda_n)\eta^{kq}
	F_{\lambda\sigma}^d+\partial_\sigma(e^\lambda_ne^\kappa_q)\eta^{kq}
	F_{\lambda\kappa}^d\right]\nonumber\\
	&&+\frac{1}{64}g^4\delta^m_nf_{ebc}f_{e'b'c'}f_d^{~b'c}f_{d'}^{~ac'}
	F_{\mu\lambda}^eF_{\mu'\lambda'}^{e'}F_{\nu\kappa}^d
	F_{\nu'\kappa'}^{d'} g^{\rho\sigma}\partial_\rho(g^{\mu'\nu'}
	x^{\lambda'}x^{\kappa'})\partial_\sigma(g^{\mu\nu}x^\lambda x^\kappa)
	\nonumber\\
	&&+\frac{1}{4}g^2\delta^m_n f_{ebc}f_d^{~ac}F_{\mu\lambda}^e
	F_{\nu\kappa}^d g^{\rho\sigma}\partial_\rho(g^{\mu\nu}x^\lambda
	x^\kappa)
	\times\partial_\sigma\left[e^\epsilon_k A^b_\phi(\partial_\epsilon
	e^\phi_b)\eta^{kp}-e^\phi_pA^b_\epsilon(\partial_\phi e^\epsilon_k)
	\right]\nonumber\\
	&&+\mbox{curvature terms}
\end{eqnarray}
Now, for the spacetimes considered by Elizalde et al., \cite{eli}, namely $
S^2\times R^2, S^1\times S^1\times R^2$, the vierbeins can be chosen to
only depend on the radii, and hence ${\cal E}_n^{mp}\equiv 0$, the only
curvature dependency then comes from the Ricci tensor. Since this is
a constant, $R_{\mu\nu}\propto \rho^{-2}$ (where $\rho$ is the radius of
$S^2$, and the result for the torus $S^1\times S^1$ is slightly more 
complicated but of the same nature), only $\cal A$ will contain this, 
while $\cal B,C$ will be curvature independent. In these
cases, furthermore, $\cal B,C$ become diagonal (except, perhaps, in colour 
space). Explicitly,
\begin{eqnarray}
	{\cal A}^{m(a)}_{n(b)} &=& -\frac{1}{2}g f_{b~c}^{~a}\eta^{mp}
	e^\mu_n e^\nu_pF_{\mu\nu}^c+\frac{1}{8}g^2\delta^m_ng^{\mu\nu}
	f_{ebc}f_d^{~ac}F_{\mu\rho}^eF_{\nu\sigma}^d x^\rho x^\sigma
	+\delta^a_bR^m_n\\
	{\cal B}^{m(a)}_{n(b)} &=& \frac{1}{8}g^2\delta^m_n f_{ebc}
	f_d^{~ac}F_{\mu\lambda}^eF_{\nu\kappa}^d g^{\rho\sigma} g^{\mu\nu}
	(\delta_\rho^\lambda\delta^\kappa_\sigma+\delta^\lambda_\sigma
	\delta^\kappa_\rho)\\
	{\cal C}^{m(a)}_{n(b)} &=& \frac{1}{64}g^4\delta^m_ng^{\rho\sigma}
	f_{e'b'c'}f_{d'}^{~ac'}f_{ebc}f_d^{~b'c}g^{\mu\nu}g^{\mu'\nu'}
	\times\nonumber\\
	&&F_{\mu'\lambda'}^{e'}F_{\nu'\kappa'}^{d'}F_{\mu\lambda}^e
	F_{\nu\kappa}^d(\delta^{\lambda'}_\rho x^{\kappa'}+
	\delta^{\kappa'}_\rho x^{\lambda'})(\delta^\lambda_\sigma x^\kappa+
	\delta^\kappa_\sigma x^\lambda)
\end{eqnarray}
With the explicit choice of the the field strength tensor used by Elizalde
et al., (the first two coordinates being $S^2$, the last two $R^2$)
\begin{displaymath}
	F_{\mu\nu}^a = \left(\begin{array}{cccc} 
		0 & 0 & 0 & 0\\
		0 & 0 & 0 & 0\\
		0 & 0 & 0 & H^a\\
		0 & 0 & -H^a & 0
	\end{array}\right)
\end{displaymath}
and using that the Ricci tensor is simply
\begin{equation}
	R_{\mu\nu} = \left(\begin{array}{cccc}
		\rho^{-2} & 0 & 0 & 0\\
		0 & 1 & 0 & 0\\
		0 & 0 & 0 & 0\\
		0 & 0 & 0 & 0
	\end{array}\right)
\end{equation}
we get $\cal A$ to be block diagonal
\begin{equation}
	{\cal A} = \left(\begin{array}{cccc}
		X+\rho^{-2} & 0 & 0 & 0\\
		0 & X+1 & 0 & 0\\
		0 & 0 & X & h\\
		0 & 0 &-h & X
	\end{array}\right) \equiv \left(\begin{array}{cc} \tilde{X} & 0\\
		0 & a \end{array}\right)
\end{equation}
where ($x,y$ are the coordinates of $R^2$)
\begin{eqnarray}
	X &=& \frac{1}{8} g^2 g^{\mu\nu} f_{ebc}f_d^{~ac}F_{\mu\rho}^e
	F_{\nu\sigma}^d x^\rho x^\sigma = \frac{1}{8}g^2 f_{ebc}f_d^{~ac}
	H^eH^d(y^2-x^2)\\
	h &=& -\frac{1}{2}g f_{b~c}^{~a}H^c
\end{eqnarray}
The logarithm of $\cal A$ appears
in the result for the effective action, since $\cal A$ is block diagonal the
calculation of this quantity is straightforward, it is
\begin{equation}
	\ln {\cal A} = \left(\begin{array}{cc} \ln \tilde{X} & 0\\
	0 & b \end{array}\right)
\end{equation}
with $a=e^b$. Now, any $2\times 2$ matrix can be expanded on the Pauli 
matrices, $a = a_0 1_2+a_i\sigma^i$, using the algebraic properties of these
we then get
\begin{equation}
	b = \left(\begin{array}{cc}
	\ln X-\ln\cos h & i h\\
	-ih & \ln X-\ln\cos h
	\end{array}\right)
\end{equation}
As for Elizalde and coworkers, the effective action gets an imaginary part
from the logarithmic term.\\
The explicit form for the remaining matrices turn out to be
\begin{eqnarray}
	{\cal B}_{n(b)}^{m(a)} &=& \delta^m_n \frac{1}{2}g^2 f_{ebc}f_d^{~ac}
	H^eH^d \equiv \delta^m_n h_2\\
	{\cal C}_{n(b)}^{m(a)} &=& \delta^m_n \frac{g^4}{16}f_{ebc}f_{e'b'c'}
	f_d^{~b'c}f_{d'}^{ac'}H^eH^{e'}H^dH^{d'}(y^2-x^2) \equiv \delta^m_n h_4
\end{eqnarray}
Inserting all of this into our general formula (\ref{eq:veff}) we then
finally arrives at
\begin{eqnarray}
	V_{\rm eff} &=& (4\pi)^{-2}{\rm tr}\left\{-\frac{g^6}{128}\left(
	(X^2+\rho^{-2})^2\ln\frac{g^2}{4}(X^2+\rho^{-2})\right.\right.\nonumber\\
	&&+(X+1)^2\ln
	\frac{g^2}{4}(X+1)+2(X^2-h^2)\ln\frac{g^2}{4}X
	\nonumber\\
	&&\left.-2(X^2-h^2)(\ln\cos\frac{g^2}{4}h+\frac{1}{2}ig^2h^2X)
	\right)
	+\frac{3g^6}{256}\left(X^2+\rho^{-2})^2+3X^2-2h^2+2X+1
	\right)\nonumber\\
	&&-\frac{1}{2}h_2\left(\ln\frac{g^2}{4}(X+\rho^{-2})+2\ln\frac{g^2}{4}X
	-2\ln\cos\frac{g^2}{4}h+\ln\frac{g^2}{4}(X+1)\right)\nonumber\\
	&&\left.-\frac{16}{3g^4}\frac{h_4}{(h^2+X^2)(X+1)(X+\rho^{-2})}
	\left((h^2+X^2)(2X+1+\rho^{-2})+2X(X+\rho^{-2})(X+1)
	\right)\right\}\nonumber\\
	&&-\frac{g^2}{2}H^aH_a
\end{eqnarray}
where the trace is over gauge algebra indices (the result is valid
for an arbitrary Lie algebra).\\
The result which Elizalde et al. finds is ($\Omega$ being the volume of
the two-sphere)
\begin{displaymath}
	\frac{\Gamma}{\Omega} = a_0 (gH)^2\left[\frac{11}{48\pi^2}
	\left(\ln\frac{gH}{\mu^{'2}}-\frac{1}{2}\right)-i\frac{1}{8\pi}\right]
	+\frac{1}{4\pi^2}\frac{gH}{\rho^2}\left[-\frac{a_0+a_1}{2}\ln 2
	+ia_1\frac{\pi}{2}\right]
\end{displaymath}
where $a_0,a_1$ are coefficients in the Schwinger-DeWitt expansion of the
heat kernel of the Laplacean on $S^2$, ($a_0=1,a_1=-1/3$), and where the gauge
group has been chosen to be $SU(2)$. A number of approximations have been made
here, first of all the calculation is linear in the curvature and secondly
it is only valid for $\rho$ large. This latter comment, however, does not 
refrain the authors of \cite{eli} from studying $1\leq \rho\leq 10$, a regime
in which the approximation $\rho \gg 1$ certainly cannot be said to hold. With
this they claim to find a critical point at $\rho_c\sim 2$. One should note
that the calculation put forward here does not suffer from these problems.\\
In our calculation we have used the freedom in renormalisation to fix $\mu=1$.
The form of the result by Elizalde and coworkers is $(H^2)(\ln H-1/2)+ R H$
with $R=\rho^{-2}$, these two terms are, up-to a finite renormalisation, the
same as the lowest order terms in our formula. The remaining terms are due to
non-linear terms in the curvature and cannot therefore be found by the
mode summation of Elizalde, Odintsov and Romeo.

\section*{Conclusion}
We have shown how the effective action for a covariantly constant
Yang-Mills field on a simple background $S^2\times R^2$ can be found, including
terms non-linear in the curvature, thereby generalising the result of Elizalde,
Odintsov and Romeo. Furthermore, as our method is not based on an explicit
mode summation, it is applicable to spacetimes in which one does not know the
explicit form of the eigenvalues of the Laplace operator for spin one (i.e.
almost all spacetimes). An important example of such a spacetime manifold, mentioned 
also by Elizalde et al., is de Sitter $S^4$. Since our approach only needs the
vierbein (and through them the Ricci tensor etc.), this spacetime is actually
within reach of the method proposed here. Manifolds with non-constant curvature
can also be treated, which means that all manifolds of physical interest should
be tractable.

Our result is also capable of handling non-covariantly
constant field configurations and to calculate curvature induced mean-fields
as shown in \cite{os,os2}, where also phase transitions are treated. 
Further research into this is in progress.

\end{document}